\documentclass[aps,reprint,amsmath,amssymb,groupedaddress,notitlepage]{revtex4-1}
\usepackage[utf8]{inputenc}
\usepackage{bm}
\usepackage{enumerate}
\usepackage{graphicx}
\usepackage{cancel}
\usepackage{xcolor}
\usepackage{verbatim}
\usepackage{hyperref}
\hypersetup{colorlinks=true,
linkcolor=blue,
citecolor=blue,
urlcolor=blue,
filecolor=blue
}

\def\bra#1{\langle{#1}|}
\def\ket#1{|{#1}\rangle}
\newcommand{\bkm}[3]{\langle #1 | #2 | #3 \rangle}
\newcommand{\rr}{{\bf r}}

\renewcommand{\d}[1]{\, \text{d} #1 \,}

\newcommand{\vk}{\mathbf{k}}
\renewcommand{\vr}{\mathbf{r}}

\newcommand{\udd}{\left| \uparrow \downarrow \downarrow \right\rangle}

\newcommand{\dud}{\left| \downarrow \uparrow \downarrow \right\rangle}
\newcommand{\ddu}{\left| \downarrow \downarrow \uparrow \right\rangle}
\newcommand{\ddd}{\left| \downarrow \downarrow \downarrow \right\rangle}

\begin{abstract}
Exchange-only spin qubits hosted in $^{28}$Si-based triple quantum dots do not suffer from decoherence caused by randomly fluctuating nuclear-spin ensembles and can be relatively robust against electrical noise when operated at a sweet spot.
Remaining sources of decoherence are qubit relaxation, leakage out of the qubit subspace, and dephasing due to residual effects of charge noise, the latter two of which are the focus of this work.
We investigate spin-orbit-mediated leakage rates to the three-spin ground state accompanied by virtual (\textit{i}) tunneling, (\textit{ii}) orbital excitation, and (\textit{iii}) valley excitation of an electron.
We find different power-law dependencies on the applied magnetic field $B$ for the three mechanisms as well as for the two leakage rates, ranging from $\propto B^5$ to $\propto B^{11}$, and identify the sweet spot as a point of minimal leakage.
We also revisit the role of electrical noise at the sweet spot, and show that it causes a decay of coherent qubit oscillations that follows a power law $\propto 1/t$ (as opposed to the more common exponential decay) and introduces a $\pi/2$ phase shift.
\end{abstract}

\begin{document}

\title{Leakage and dephasing in $^{28}$Si-based exchange-only spin qubits}

\author{Arnau Sala}
\affiliation{Center for Quantum Spintronics, Department of Physics, Norwegian University of Science and Technology, NO-7491 Trondheim, Norway}

\author{Jeroen Danon}
\affiliation{Center for Quantum Spintronics, Department of Physics, Norwegian University of Science and Technology, NO-7491 Trondheim, Norway}

\date{\today}

\maketitle


The development of semiconductor quantum-dot spin qubits seems to be a promising path towards the materialization of large-scale quantum computation~\cite{Loss1998}.
In order to overcome the practical challenge of creating highly localized oscillating magnetic fields, implementations of such qubits have seen a development from single-dot single-spin systems to a more complicated triple-dot three-spin exchange-only (XO) qubit that can be fully operated by electric fields only~\cite{Petta2005,Koppens2006,DiVincenzo2000,Taylor2013,Medford2013a,Malinowski2017}.
Furthermore, hosting spin qubits in purified $^{28}$Si, instead of the more traditional III-V materials, led to a significant improvement of observed qubit coherence times, due to the negligible fraction of spinful nuclei in the material~\cite{Medford2013,Simmons2009,Zwanenburg2013,Veldhorst2014,Muhonen2014,Eng2015,Martins2016}.

Remaining sources of decoherence for the $^{28}$Si-based XO qubit are
(\textit{i}) electric noise in the environment of the qubit leading to qubit dephasing~\cite{Russ2017,Zhang2018},
(\textit{ii}) electron-phonon coupling that can cause (spin-conserving) qubit relaxation~\cite{Taylor2013}, and
(\textit{iii}) spin-mixing mechanisms such as spin-orbit (SO) interaction that can enable leakage out of the qubit subspace to the three-spin ground state $\ddd$~\cite{Hanson2007,Russ2017}.
Some of these mechanisms have already been studied:
It was found that the effects of charge noise can be strongly suppressed by manipulating the qubit at a so-called sweet spot (SS), where the qubit splitting is to leading order insensitive to electric fluctuations~\cite{Shim2016,Malinowski2017}, and electron-phonon coupling was shown to cause slow qubit relaxation (estimated as $\Gamma_\text{rel} \lesssim 10$~Hz) that is proportional to the fifth power of the qubit splitting~\cite{Srinivasa2016};
some effects of SO interaction can also be suppressed during gate operations in double quantum dots by shaping the pulse of the two-qubit coupling~\cite{Bonesteel2001,Burkard2002} or by using superexchange couping in a triple quantum dot setup~\cite{Rancic2017}.

In this work we study some of the remaining questions.
We first investigate the SO-induced leakage rates from the two qubit states to the ground state $\ddd$.
Since a SO-assisted spin flip requires finite motion of the electron, such a leakage process must involve virtual excitation of a different orbital state~\cite{Hanson2007}; here we consider the contributions from virtual tunneling, on-site orbital excitation, and valley excitation separately.
For these three mechanisms we find different power laws for the dependence of the two rates on the applied magnetic field $B$, ranging from $\Gamma_\text{leak} \propto B^5$ to $\Gamma_\text{leak} \propto B^{11}$, and we also show that the SS is the point where both the qubit relaxation and leakage rates are minimal.
Finally, we also revisit the role of charge noise at the SS and we show that slow electric fluctuations in the qubit's environment cause a power-law decay $\propto 1/t$ of coherent qubit oscillations, as opposed to the exponential decay that is usually assumed~\cite{Russ2015,Russ2016}.

The rest of this paper is organized as follows:
In Sec.~\ref{sec:mod} we introduce our description of the system and the model Hamiltonians we use.
In Sec.~\ref{sec:anal} we present our analytic results for the leakage rates, based on the three mechanisms mentioned above.
Then, in Sec.~\ref{sec:num}, we corroborate these results with a numerical evaluation of the dominating leakage rates, across the whole (1,1,1) charge region.
In Sec.~\ref{sec:dep} we investigate charge-noise-induced dephasing, and in Sec.~\ref{sec:conc} we finally present our conclusions.

\section{Model}
\label{sec:mod}

\begin{figure}
\includegraphics[width=\linewidth]{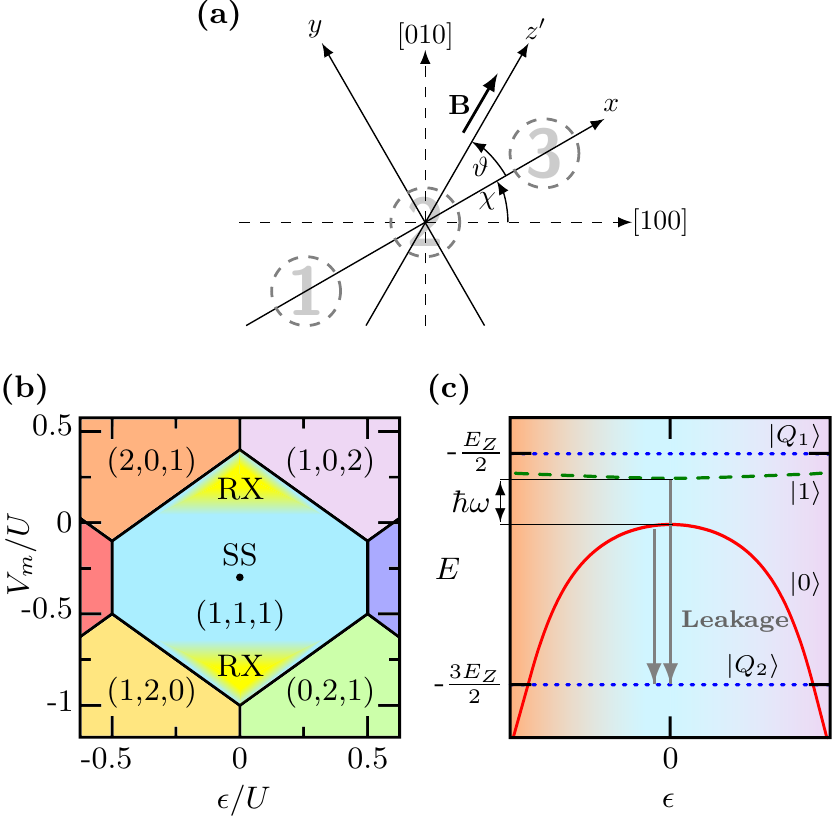}
\caption{(a) Reference frame of the system. The quantum dots (dashed gray circles) lie in the $xy$-plane. An in-plane magnetic (Zeeman) field $\mathbf{B}$ is applied at an angle $\vartheta$ with the interdot axis, which, in turn, is at an angle $\chi$ with the crystallographic [100] direction. (b) Charge stability diagram, showing the charge ground state of the electrostatic part of the Hamiltonian~(\ref{eq:hubb}) as a function of $\epsilon$ and $V_m$, using $U_c = 0.3\, U$ and $V_2 = 2U_c-V_1-V_3$. (c) Qualitative sketch of the lower part of the spectrum of (\ref{eq:hubb}) as a function of $\epsilon$ in the regions labeled `RX' in (b), where we assumed $t_{12}=t_{23}$.
The gray arrows indicate the leakage processes investigated here. In this plot all variables are in arbitrary units. \label{fig:ref-frame}}
\end{figure}

We consider a linear array of three circular quantum dots with radius $\sigma$ and interdot distance $d$ (center-to-center), as schematically depicted in Fig.~\ref{fig:ref-frame}a.
Assuming a large orbital level splitting on the dots, we allow each dot $i$ to contain $n_i\in\{0,1,2\}$ excess electrons, and the triplet $(n_1,n_2,n_3)$ will hereafter be used to label the different charge configurations.
We model the system using a Hubbard-like Hamiltonian~\cite{Medford2013a,Taylor2013,DasSarma2011}
\begin{align}\label{eq:hubb}
\hat H = {}&{} \sum_i \left[ \frac{U}{2}\hat n_i(\hat n_i - 1) - V_i \hat n_i \right] + \sum_{\langle i,j\rangle} U_c \hat n_i \hat n_j \notag \\
{}&{} 
 + \sum_{\langle i,j\rangle, \alpha} \frac{t_{ij}}{\sqrt 2} \hat c_{i,\alpha}^\dagger \hat c_{j,\alpha}
 + \sum_{i,\alpha} \frac{1}{2}g \mu_B B \hat c_{i,\alpha}^\dagger \sigma_{z'}^{\alpha \alpha}\hat c_{i,\alpha},
\end{align}
where $\hat n_i = \sum_\alpha \hat c_{i,\alpha}^\dagger \hat c_{i,\alpha}$, with $\hat c_{i,\alpha}^\dagger$ the creation operator for an electron with spin $\alpha$ in the orbital ground state of dot $i$, and $\hat\sigma_{z'}$ is the diagonal Pauli matrix, acting in spin space.
As in Refs.~\cite{Taylor2013,Sala2017}, the first line describes the electrostatic energy and includes an on-site charging energy $U$, gate-tunable local potentials $V_i$, and a nearest-neighbor charging energy $U_c$.
The second line adds nearest-neighbor (spin-conserving) interdot tunnel couplings and a Zeeman splitting due to an externally applied magnetic field $B$, which we assume to be in-plane.
The tunnel coupling parameters could be effectively renormalized due to phase differences between the valley states on neighboring dots~\cite{Note0}; we assume such effects to be included in the $t_{ij}$ we use.

Fig.~\ref{fig:ref-frame}b shows part of the charge stability diagram resulting from the first line of (\ref{eq:hubb}), where the regions with different charge (ground) states are indicated, as a function of $\epsilon=(V_3-V_1)/2$ and $V_m = (V_1+V_3)/2-V_2$ for $U_c = 0.3\, U$ and $V_1+V_2+V_3 = 2U_c$.
Within the (1,1,1) region finite tunnel couplings $t_{ij}$ lead to exchange effects that split the spectrum in a fourfold degenerate spin quadruplet and two doubly degenerate doublets.
The additional Zeeman field $B$ further splits all states with different total spin projection $S_z^\text{tot}$, and in Fig.~\ref{fig:ref-frame}c we qualitatively sketch the resulting lowest part of the spectrum of (\ref{eq:hubb}) in the (1,1,1) region as a function of $\epsilon$,
where we assume that $t_{12} = t_{23} \equiv \tau$ and use $E_\text{Z} = g\mu_B B$.

At $\epsilon=0$ the two spin doublet states with $S^\text{tot}_z = -\frac{1}{2}$ are $\ket{0} = ( \ddu + \udd - 2\dud )/\sqrt{6}$ and $\ket{1} = ( \ddu - \udd )/\sqrt{2}$ and provide a basis for a qubit that can be controlled fully by electrical means~\cite{Laird2010,Medford2013a,Taylor2013,Russ2017,Malinowski2017,Medford2013}:
The qubit splitting reads (to lowest order in $\tau$) $\hbar\omega = 2\tau^2 (U-U_c)/E_oE_i$, with $E_o = U-2U_c-V_m$ and $E_i = U+V_m$, which can be controlled through $\tau$ and $V_m$, and a small $\epsilon$ yields a term $\propto \epsilon \, \hat\sigma_x$ in the projected qubit Hamiltonian~\cite{Note1}.
Close to the borders of the (1,1,1) region (the regimes labeled `RX' in Fig.~\ref{fig:ref-frame}b) a small modulation of $\epsilon$ with frequency $\omega$ thus induces Rabi oscillations.
This so-called resonant-exchange (RX) regime has the advantage that the qubit operations can be fast~\cite{Medford2013,Medford2013a}.
At the center of the (1,1,1) region (the ``sweet spot'' labeled `SS') the qubit should be operated with larger pulses (resonant or static)~\cite{Shim2016,Malinowski2017,Martins2016}, but here one has the benefit that the qubit splitting is to leading order insensitive to noise in the gate potentials.
The qubit dephasing time $T_2^*$ is thus predicted to be orders of magnitude larger at this point than in the RX regime~\cite{Huang2014,Russ2015}.
Below we will investigate the remaining dephasing at the SS in more detail.

The leading effects of charge noise can thus be suppressed by operating the qubit at the SS and, since $^{28}$Si is nuclear-spin-free, the hyperfine interaction that reduces the dephasing time in GaAs-based spin qubits to $\sim 10$~ns~\cite{Medford2013,Gaudreau2011,Medford2013a,Peterfalvi2017} is not a concern here.
That leaves as possibly dominating decoherence mechanisms (\textit{i}) qubit relaxation (transitions from $\ket{1}$ to $\ket{0}$) due to electron-phonon coupling and (\textit{ii}) leakage out of the qubit space (dissipative transitions to the ground state $\ket{Q_2} = \ddd$) enabled by SO interaction combined with electron-phonon coupling, see the gray arrows in Fig.~\ref{fig:ref-frame}c~\cite{Khaetskii2000,Khaetskii2001,Maisi2016}.
Since phonon-mediated relaxation of the triple-dot EO qubit has been studied before~\cite{Taylor2013,Srinivasa2016,Russ2017}, we will focus here on the leakage caused by SO interaction.

We model the SO coupling for each electron with the Hamiltonian~\cite{Tahan2005,Hanson2007}
\begin{align} \label{eq:hamsoi}
	\hat H_\text{SO} = A_{xx} \hat p_x \hat\sigma_{x'}
	+ A_{yx} \hat p_y \hat\sigma_{x'},
\end{align}
where $\hat{\bf p}$ is the electron's momentum.
We exclusively focus on the spin-flip terms $\propto \sigma_{x'}$ (see Fig.~\ref{fig:ref-frame}a) and used $A_{xx} = \alpha \cos\vartheta + \beta ( \cos\vartheta \sin 2\chi + \sin\vartheta \cos 2\chi )$ and $A_{yx} = \alpha \sin\vartheta + \beta ( \cos\vartheta \cos 2\chi - \sin\vartheta \sin 2\chi )$, where $\alpha$ and $\beta$ are the amplitudes of the Rashba and Dresselhaus terms, respectively.
Rashba SO coupling in Si-based quantum wells is predicted to come from structural inversion asymmetry arising from electric fields set up by interface effects~\cite{Tahan2014,Jock2018}.
Dresselhaus SO coupling is usually associated with inversion asymmetry of the crystal lattice, which is in principle absent in Si~\cite{Zwanenburg2013,Dresselhaus1955}.
However, theoretical work predicted that microscopic details (such as the exact number of atomic Si layers in the well or roughness of the interfaces) can give rise to a Dresselhaus-like term that could be comparable to or even dominate over the Rashba term~\cite{Prada2011,Golub2004,Nestoklon2008}; this was recently confirmed by several experiments~\cite{Ferdous2018,Jock2018,Tanttu2018}.

For the electron-phonon coupling we use the Hamiltonian~\cite{Danon2013,Tahan2014,Malkoc2016}
\begin{align} \label{eq:eph}
	\hat{H}_\text{e-ph} = \sum_{\vk,p} \lambda_{\vk,p} \hat{\rho}_\vk \left( \hat{a}_{\vk,p} + \hat{a}^\dagger_{-\vk,p} \right),
\end{align}
with $\hat{\rho}_\vk = \int \d\vr e^{-i\vk\cdot \vr} \hat{\rho}(\vr)$ the Fourier transform of the electronic density operator, $\hat a^\dagger_{\vk,p}$ the creation operator of a phonon with wave vector $\vk$ and polarization $p$, and the coupling parameters $\lambda_{\vk,p}$ given by
\begin{align}
	|\lambda_{\vk,p}|^2 = {}&{} 
	\frac{\hbar}{2\rho_0 v_p k \mathcal{V}} \left[ \Xi_d {\bf e}_k^p\cdot \vk + \Xi_u ({\bf e}_k^p)_z k_z \right]^2,
\end{align}
with $v_p$ the (polarization-dependent) sound velocity, $\rho_0$ the electron density, ${\cal V}$ the normalization volume, $\Xi_u$ and $\Xi_d$ the uniaxial shear and dilatation deformation potentials~\cite{Herring1956},
and ${\bf e}_k^l$ and ${\bf e}_k^t$ unit vectors along the longitudinal and transversal directions of phonon propagation.

For localized electrons, such as in quantum dots, $\langle \hat{\bf p}\rangle$ vanishes and therefore SO interaction does not directly couple states that have the same orbital wave function but opposite spin.
Spin-flip transitions within the orbital ground state thus require the excitation of a virtual state which involves finite motion of the electrons.
We will investigate three such mechanisms:
(\textit{i}) virtual tunneling to a neighboring dot (an ``exchange-enabled'' spin flip); 
(\textit{ii}) virtual excitation of a higher orbital on the same dot;
and (\textit{iii}) virtual excitation of the other valley state, also on the same dot.
All three mechanisms, in combination with the emission of a phonon to ensure energy conservation, can thus lead to a spin flip and thereby cause leakage out of the qubit space as discussed above.

We should note, however, that in realistic systems one often cannot treat the orbital or valley index of an excited state as a good quantum number, the actual states being of a mixed valley-orbital nature~\cite{Zwanenburg2013,Friesen2010}.
For clarity of presentation, we will first investigate the cases of pure virtual orbital and pure virtual valley excitation separately, and then, at the end of Sec.~\ref{sec:val}, discuss how our results relate to the case of mixed valley-orbital states.

\section{Analytic results}
\label{sec:anal}

We now investigate the three spin-orbit-mediated leakage mechanisms in more detail, and we evaluate the leakage rates $\Gamma_{1,0}$ from qubit states $\ket 1$ and $\ket 0$ to the ground state $\ket{Q_2}$, focusing on $\epsilon = 0$ and symmetric tunnel coupling, $t_{12} = t_{23} \equiv \tau$.
In all cases we will calculate the rates using a second-order Fermi's golden rule,
\begin{align}
	\Gamma_\alpha =  \sum_{{\bf k},p} \frac{2\pi}{\hbar} \left| \sum_v \frac{\bra{f} \hat H' \ket{v}\bra{v} \hat H' \ket{i}}{E_v - E_i} \right|^2 \delta ( E_f - E_i),
\end{align}
where $\hat H' = \hat H_{\rm SO} +  \hat H_\text{e-ph}$ and the second sum runs over all possible virtual states $\ket v$.
The initial state $\ket i$ is $\ket{\alpha; {\rm vac}}$ with $\alpha \in \{ 1,0\}$ (one of the qubit states combined with the phonon vacuum) and the final state $\ket f$ is $\ket{Q_2; 1_{{\bf k},p}}$ (the ground state combined with one phonon with wave vector ${\bf k}$ and polarization $p$).

\subsection{Virtual spin-flip tunneling}
\label{sec:ex}

By using that $\hat{\mathbf{p}} = \frac{i}{\hbar}m^* [\hat H,\hat{\mathbf{r}}]$, with $\hat H$ as in (\ref{eq:hubb}) and $m^*$ the effective electron mass, one can derive matrix elements of $\hat H_{\rm SO}$ that couple states with different spin and charge configurations~\cite{Danon2013,Golovach2008}.
These ``spin-flip tunneling'' matrix elements couple both $\ket 1$ and $\ket 0$ to $\ket{Q_2}$ and phonon emission is then governed by matrix elements that do not alter the spin or orbital state of the electrons.
Assuming a parabolic confinement in the quantum dots, and thus a Gaussian ground state envelope wave function, the matrix element describing the emission of a phonon by an electron in quantum dot $j$ at position $x_j$ reads
\begin{align} \label{eq:eph-ex}
{} & {} \bkm{0_j{-}s;1_{{\bf k},p}}{\hat H_\text{e-ph}}{0_j{-}s;{\rm vac}} = \\ 
{}&{} \ \ i \sqrt{\frac{\hbar}{2\rho_0 v_p k \mathcal{V}}} \left[ \Xi_d {\bf e}_k^p\cdot \vk + \Xi_u ({\bf e}_k^p)_z k_z  \right] 
e^{-\frac{1}{4}(k_x^2+k_y^2)\sigma^2-i k_x x_j},\notag 
\end{align}
where $\ket{0_j{-}s}$ denotes the state of an electron in the ground state in dot $j$ with spin $s$ and in the lowest valley state (denoted by `$-$').

Taking into account all three electrons, we can arrive at analytic expressions for the leakage rates.
Defining $\Delta^{-2} = E_o^{-2} - E_i^{-2}$, we find to leading order in $\tau/\Delta$
\begin{align}
	\Gamma_{1} \approx {}&{}
	\frac{1}{16\pi}
	\frac{d^2}{l_{\rm so}^2} 
	\frac{\tau^4}{\Delta^4}
	\frac{\Xi_u^2E_\text{Z}^3}{\hbar^{4}v_t^5 \rho_0}
	\, f^\text{ex}_1\left( \frac{dE_\text{Z}}{\hbar v_t}\right), \\
	\Gamma_{0} \approx {}&{}
	\frac{3}{16\pi}
	\frac{d^2}{l_{\rm so}^2} 
	\frac{\tau^4}{\Delta^4}
	\frac{\Xi_u^2 E_\text{Z}^3}{\hbar^4 v_t^5 \rho_0}
	\, f^\text{ex}_0\left( \frac{dE_\text{Z}}{\hbar v_t}\right),
\end{align}
where $l_{\rm so} = \hbar / m^* A_{xx}$ is the relevant spin-orbit length and the functions $f^\text{ex}_{1,0}(x) \sim 1$ for $x \gtrsim 1$; they are given explicitly in Appendix~\ref{appA}.
To arrive at these expressions, we assumed that $E_\text{Z} \gg \tau^2/E_{i,o}$ (which is typically satisfied if $B \gtrsim 10$~mT) and $E_\text{Z}^2 \ll (\hbar v_t / \sigma)^2$ (which, for $\sigma = 15$~nm, limits $B \lesssim 2$~T).
Furthermore, we used that in Si $v_l \approx 2 v_t$, which makes $(v_t/v_l)^5 \ll 1$.
For small $E^2_Z \ll (\hbar v_t/d)^2$ we can expand the functions $f^\text{ex}_{1,0}(x)$ in small $x$, yielding
\begin{align}
	\Gamma_{1} \approx {}&{}
	\frac{1}{840\pi}
	\frac{d^2}{l_{\rm so}^2} 
	\frac{\tau^4}{\Delta^4}
	\frac{d^4\Xi_u^2E_\text{Z}^7}{\hbar^{8}v_t^9 \rho_0}, \label {eq:rateRX1}\\
	\Gamma_{0} \approx {}&{}
	\frac{1}{35\pi}
	\frac{d^2}{l_{\rm so}^2}
	\frac{\tau^4}{\Delta^4}
	\frac{d^2 \Xi_u^2 E_\text{Z}^5}{\hbar^6 v_t^7 \rho_0}. \label {eq:rateRX0}
\end{align}
We see that $\Gamma_1 \propto E_\text{Z}^7$ and $\Gamma_0 \propto E_\text{Z}^5$ in this limit, and $\Gamma_1$ is smaller than $\Gamma_0$ by a factor $(dE_\text{Z}/\hbar v_t)^2$. 
In case of substantially asymmetric tunneling amplitudes (i.e., $t_{12}\neq t_{23}$) we find that both rates scale as $\Gamma \propto E_\text{Z}^5$.

These rates can be directly compared with the qubit relaxation rate, from $\ket 1$ to $\ket 0$, which comes mainly via electron-phonon coupling~\cite{Taylor2013,Khaetskii2001}
\begin{align}
\Gamma_{\rm rel} = {}&{}
\frac{1}{70\pi}
\frac{\tau^4}{\Delta^4}
\frac{d^2\Xi_u^2(\hbar\omega)^5}{\hbar^6v_t^7\rho_0},
\end{align}
where $\hbar \omega$ is the qubit splitting.

Note that all relaxation rates cancel when $1/\Delta=0$, which happens at the SS.
A qubit operated at this point will therefore not only be highly insensitive to both dephasing due to charge noise, but also to relaxation and leakage.
Additionally, the dependence of $A_{xx}$ on the angles $\chi$ and $\vartheta$ allows for a reduction of the leakage rates by varying the device orientation and the direction of ${\bf B}$: In fact, for $\chi$ equal to a multiple of $\pi/2$, we see that there are angles $\vartheta = (n+\frac{1}{2})\pi$ for which $A_{xx} = 0$.
Such strong angular dependence of spin-orbit-mediated relaxation rates is already well known from theory and experiments on double quantum dots~\cite{Hofmann2017,Raith2012,Raith2012a,Golovach2004}.

\subsection{Virtual orbital excitation}
\label{sec:orb}

The parabolic potential that confines the electrons in the quantum dots results in Fock-Darwin eigenstates with energy splitting $E_\text{orb} = \hbar^2/m^* \sigma^2$.
The SO interaction couples the orbital ground state to the first excited state with opposite spin~\cite{Hanson2007},
\begin{align}
	\bkm{1^\alpha_j{-}\bar s}{\hat p_\alpha \hat \sigma_{x'}}{0_j{-}s} = 
\frac{i\hbar}{\sqrt{2}\sigma},
\end{align}
where $\bar s$ denotes the spin state opposite to $s$ and the superscript $\alpha$ indicates which component of the wave function is in the excited state~\cite{Note3}.
The electron-phonon Hamiltonian also couples these two orbital states~\cite{Tahan2014},
\begin{align}
{} & {} \bkm{1^\alpha_j{-}s;1_{{\bf k},p}}{\hat H_\text{e-ph}}{0_j{-}s;{\rm vac}} = \notag \\ 
 {}&{} \hspace{2em} i \sqrt{\frac{\hbar}{2\rho_0 v_p k \mathcal{V}}} [ \Xi_d {\bf e}_k^p\cdot \vk + \Xi_u ({\bf e}_k^p)_z k_z  ] g_{10}^{(j,\alpha)}(\vk),
\end{align}
with $g_{10}^{(j,\alpha)}(\vk)$ being the Fourier transform of the overlap between 
the ground and first excited state on dot $j$,
\begin{align}
g_{10}^{(j,\alpha)}(\vk) = {}&{} -\frac{i}{\sqrt{2}} k_\alpha \sigma e^{-\frac{1}{4}(k_x^2+k_y^2) \sigma^2 -i k_x x_j}.
\end{align}

The resulting leakage rates, involving the virtual excitation of an orbital state, can straightforwardly be evaluated.
Compared to the exchange-enabled rates, they come with large powers of $E_\text{Z}/E_\text{orb}$ instead of $\tau/\Delta$, which makes them typically much smaller.
Under the same assumptions as before we find
\begin{align}
\Gamma_1 \approx {}&{}
      \frac{1}{4 \pi}
      \frac{E_{Z}^4}{E_\text{orb}^4}
      \frac{\Xi_u^2E_{Z}^3}{\hbar^4v_t^5\rho_0 }
      \, f_1^\text{orb}\left(\frac{d E_{Z}}{\hbar v_t}\right), \label{eq:rateORB1a}\\
\Gamma_0 \approx {}&{}
      \frac{1}{12 \pi}
      \frac{E_{Z}^4}{E_\text{orb}^4}
      \frac{\Xi_u^2E^3_{Z}}{\hbar^4 v^5_t\rho_0}
      \, f_0^\text{orb}\left(\frac{d E_{Z}}{\hbar v_t}\right), \label{eq:rateORB0a}
\end{align}
where the dimensionless functions $f_{1,0}^\text{orb}(x)$ are given in Appendix~\ref{appA}.
For $x \gtrsim 1$ they are of the order of $(A_{xx}^2+A_{yx}^2)/v_t^2$ and for small $E^2_Z \ll (\hbar v_t/d)^2$ we can again expand the functions, yielding
\begin{align}
\Gamma_1 \approx {}&{}
      \frac{2}{315 \pi}
      \frac{3A_{xx}^2 + A_{yx}^2}{v_t^2}
      \frac{E_\text{Z}^4}{E_\text{orb}^4}
      \frac{d^2 \Xi_u^2E_{Z}^5}{\hbar^6 v_t^7 \rho_0}, \label{eq:rateORB1} \\
\Gamma_0 \approx {}&{}
      \frac{2}{10395 \pi}
      \frac{5A_{xx}^2 + A_{yx}^2}{v_t^2}
      \frac{E^4_{Z}}{E_\text{orb}^4}
      \frac{d^4 \Xi_u^2E_{Z}^7}{\hbar^8 v^9_t\rho_0}. \label{eq:rateORB0}
\end{align}
In this case we thus find that $\Gamma_1 \propto E_\text{Z}^9$ and $\Gamma_0 \propto E_\text{Z}^{11}$ and that now $\Gamma_1$ is \textit{larger} than $\Gamma_0$ by a factor $(dE_\text{Z}/\hbar v_t)^{-2}$ (on top of a rather large difference in numerical prefactors), opposite to the exchange-enabled rates.
Comparing the two mechanisms qualitatively, we see that the factor $d^2\tau^4/l_{\rm so}^2 \Delta^4$ in the exchange-enabled rates is replaced here by a factor $A^2 E_\text{Z}^4/v_t^2 E_\text{orb}^4$, which is typically much smaller~\cite{Note4}.
Another qualitative difference is that the ``orbital-assisted'' rates (\ref{eq:rateORB1a},\ref{eq:rateORB0a}) do not depend on the tuning through $\Delta$ and thus survive at the SS.

We can compare these results with Eq.~(12) in Ref.~\cite{Tahan2014}, where the authors calculated the ground state spin relaxation rate in a \textit{single} quantum dot via virtual excitation of an orbital state.
We see that our results are fundamentally the same, apart from extra factors of $(dE_{\rm Z}/v_t\hbar)^2$, which result from the
multi-electron/multi-dot nature of our system and account for interference between spin-flip amplitudes on different dots.
If we would make the orbital energy splitting substantially different on each dot, we would also find $\Gamma \propto E_\text{Z}^7$ for both relaxation rates.

\subsection{Virtual valley excitation}
\label{sec:val}

The band gap in bulk Si is indirect and the conduction band has six minima, away from $k=0$.
In most Si-based heterostructures strain splits off four of these minima, leaving two minima at ${\bf k} \approx \pm 0.85\, k_\text{max} \hat z$, where $\hat z$ is the growth direction of the structure.
Localized electrons in the conduction band thus have an extra ``valley'' degree of freedom and can be described by a wave function
\begin{align}
\psi^{(v)} = F^{(v)}(\rr) \big[ \alpha^{(v)}_1 u_1(\rr)e^{ik_zz} + \alpha^{(v)}_2 u_2(\rr) e^{-ik_zz} \big],
\end{align}
where $F^{(v)}(\rr)$ is the envelope wave function corresponding to valley $v$ and $u_{1,2}(\rr)$ is the lattice-periodic part of the Bloch functions at the conduction band minima at $\pm k_z$.
Inhomogeneities such as disorder and interface roughness typically couple the two minima, resulting in eigenstates with $\alpha^{(\pm)}_1 = \frac{1}{\sqrt 2}$ and $\alpha^{(\pm)}_2 = \pm \frac{1}{\sqrt 2}$.

Both SO and electron-phonon interaction can couple opposite valley states~\cite{Tahan2014}, and virtual valley excitation can thus cause leakage in a way similar to virtual orbital excitation.
The relevant matrix elements, however, depend sensitively on details of the confinement along the $z$-direction that are hard to predict.
We thus take a slightly more qualitative approach and start by employing the dipole approximation $e^{-i{\bf k}\cdot{\bf r}} \approx 1 - i{\bf k}\cdot{\bf r}$ in the electron-phonon Hamiltonian (\ref{eq:eph}), which amounts to assuming that the emitted phonon has a wave length much larger than the electronic confinement length (equivalent to the assumption $E_\text{Z}^2 \ll (\hbar v_t/\sigma)^2$ used before).
This allows us to write
\begin{align}\label{eq:val}
{}&{} \bkm{0_j{+}s;1_{{\bf k},p}}{\hat H_\text{e-ph}}{0_j{-}s;{\rm vac}} \approx \\ 
{}&{} \ \ \sqrt{\frac{\hbar}{2\rho_0 v_p k \mathcal{V}}} [ \Xi_d {\bf e}_k^p\cdot \vk + \Xi_u ({\bf e}_k^p)_z k_z ]  e^{-i k_x x_j} \mathbf k \cdot \mathbf{r}_{+-}, \notag
\end{align}
with $\mathbf{r}_{+-} = \bkm{0_j{+}s}{\mathbf{r}}{0_j{-}s}$ the valley dipole matrix element.
If we use again that $\hat{\mathbf p} = \frac{i}{\hbar}m^* [ \hat H,\mathbf r ]$,
then we can express the SO Hamiltonian in terms of the same dipole matrix elements.
The precise magnitude of these elements depends again on microscopic details, and for simplicity we will use that $|z_{+-}| \ll |x_{+-}|, |y_{+-}|$ and assume $x_{+-} = y_{+-} \equiv r_d$~\cite{Huang2014a,Yang2013}.
This phenomenological parameter can be related to the magnitude of SO-induced anticrossings in the electronic spectrum between states with different spin and valley index; for Si MOS-based quantum dots $|r_d|\sim$~1--2~nm has been reported~\cite{Yang2013}.

We can now calculate the leakage rates and find using again the same assumptions
\begin{align}
\Gamma_{1} \approx {}&{}
      \frac{1}{\pi}
      \frac{A^2}{v^2_t}
      \frac{|r_d|^4}{l_Z^4}
      \frac{E_{Z}^2}{E_{v}^2}
      \frac{\Xi_u^2E^3_{Z}}{\hbar^4 v_t^5 \rho_0}
      \, f_1^\text{val}\left(\frac{d E_\text{Z}}{\hbar v_t}\right), \label{eq:rel_v1}\\
\Gamma_{0} \approx {}&{}
      \frac{1}{3 \pi}
      \frac{A^2}{v^2_t}
      \frac{|r_d|^4}{l_Z^4}
      \frac{E_{Z}^2}{E_{v}^2}
      \frac{\Xi_u^2E^3_{Z}}{\hbar^4 v_t^5 \rho_0}
      \, f_0^\text{val}\left(\frac{d E_\text{Z}}{\hbar v_t}\right), \label{eq:rel_v0}
\end{align}
where, for convenience of notation, we introduced the Zeeman length $l_Z = \hbar / \sqrt{m^* E_\text{Z}}$ and $E_v$ denotes the splitting between the two valley states.
The parameter $A \sim \alpha,\beta$ sets the strength of the SO interaction; we cannot resolve the detailed dependence on the angles $\vartheta,\chi$ in this case since that would require knowing the exact relative magnitude and phase of $x_{+-}$ and $y_{+-}$ as well.
The dimensionless functions $f_{1,0}^\text{val}(x)$, given in Appendix~\ref{appA}, are again of the order 1 for $x\gtrsim 1$ and can be expanded in small $x$ when $E^2_{Z} \ll (\hbar v_t /d)^2$, giving
\begin{align}
\Gamma_{1} \approx {}&{}
      \frac{32}{315 \pi}
      \frac{A^2}{v_t^2}
      \frac{|r_d|^4}{l_Z^4}
      \frac{ E_{Z}^2}{E_{v}^2}
      \frac{d^2 \Xi_u^2 E_\text{Z}^5}{\hbar^6 v_t^7 \rho_0}, \label{eq:rateVAL1} \\
\Gamma_{0} \approx {}&{}
      \frac{16}{3465 \pi}
      \frac{A^2}{v_t^2} 
      \frac{|r_d|^4}{l_Z^4} 
      \frac{E_{Z}^2}{E_{v}^2}
      \frac{d^4\Xi_u^2 E_\text{Z}^7}{\hbar^8 v_t^9 \rho_0}. \label{eq:rateVAL0}
\end{align}
We find again $\Gamma_1 \propto E_\text{Z}^9$ and $\Gamma_0 \propto E_\text{Z}^{11}$, as well as that $\Gamma_1$ is larger than $\Gamma_0$ by a factor $(dE_\text{Z}/\hbar v_t)^{-2}$ and that the rates do not depend on tuning parameters, all qualitatively similar to the rates based on virtual orbital excitation. 
Comparing the rest of the expressions, we find that the valley-assisted rates are smaller than the orbital-assisted ones by a factor $\sim |r_d|^4E^2_\text{orb}/\sigma^4 E_v^2$, where typically $|r_d| \sim$~1--2~nm and $\sigma \sim$~10--30~nm, which makes this a very small factor.
A significant variation of $E_v$ or $|r_d|$ over the dots would yield relaxation rates that scale as $\Gamma \propto E_\text{Z}^7$ in both cases.

In the presence of valley-orbital mixing of the excited states it is also hard to write analytic expressions for the dipole matrix elements needed.
In this case Eqs.~(\ref{eq:rel_v1}--\ref{eq:rateVAL0}) are the most useful results, where $r_d$ now describes the dipole matrix element between the ground state and first excited valley-orbital state and $E_v$ should of course be replaced by the valley-orbital ground state gap $E_{vo}$.

\section{Numerical results}
\label{sec:num}

We corroborate our approximate analytic results with a numerical evaluation of the leakage rates across the whole (1,1,1) charge region.
We focus here on the dominating exchange-assisted mechanism of Sec.~\ref{sec:ex}, which is also the only one that shows a dependence on the tuning parameters $\epsilon$ and $V_m$.

We start by diagonalizing the Hamiltonian $\hat H + \hat H_\text{SO}$, disregarding the excited orbital and valley states.
We then identify in the spectrum the two qubit states $\ket 1$, $\ket 0$ (the spin doublet states with $S_z^\text{tot} = -\frac{1}{2}$) and the quadruplet state $\ket{Q_2}$.
Using Fermi's golden rule,
\begin{align}
\Gamma_{\alpha} = {}&{} \frac{2\pi}{\hbar} \sum_{{\bf k},p} | \langle Q_2;1_{\vk,p} | \hat H_{\text{e-ph}} | \alpha;\text{vac} \rangle |^2\, \delta(E_f-E_i),
\end{align}
we finally calculate the two leakage rates numerically.

\begin{figure}
\includegraphics[width=\linewidth]{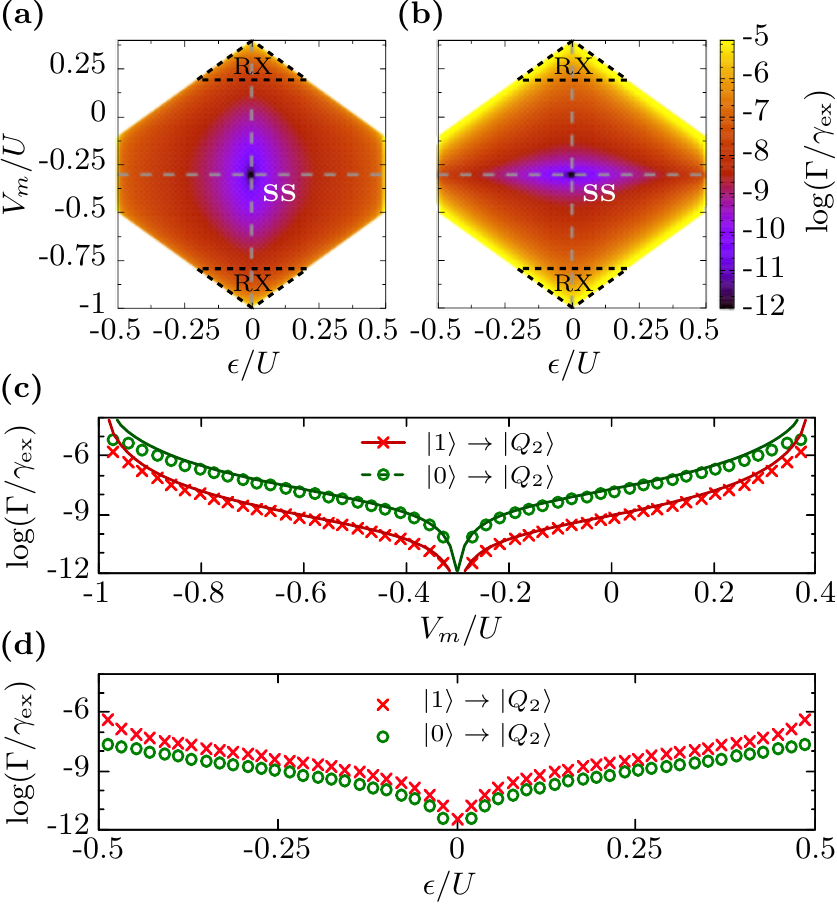}
\caption{Spin-flip-tunneling-assisted leakage rates out of the qubit space across the whole (1,1,1) charge region, from (a) $\ket 1$ and (b) $\ket 0$ to $\ket{Q_2}$ in units of $\gamma_\text{ex} \equiv d^2\Xi_u^2 E_\text{Z}^3 / l_\text{so}^2 \hbar^4v_t^5\rho_0$ (see main text for the choice of parameters).
(c) $\Gamma_{1,0}$ as a function of $V_m$ for $\epsilon = 0$ (circles and crosses), i.e., along the vertical dashed lines in (a,b).
Solid lines show the analytical results from Eqs.~(\ref{eq:rateRX1},\ref{eq:rateRX0}).
(d) $\Gamma_{1,0}$ as a function of $\epsilon$ for $V_m = -0.3\,U$, i.e., along the horizontal dashed lines in (a,b).
\label{fig:relax_rates}}
\end{figure}

The results are shown in Fig.~\ref{fig:relax_rates}, where we plot the leakage rates in units of $\gamma_\text{ex} \equiv d^2\Xi_u^2 E_\text{Z}^3 / l_\text{so}^2 \hbar^4v_t^5\rho_0$.
We used $t_{12}=t_{23} \equiv \tau=16~\mu$eV, $U=50\, \tau$, $U_c=15\, \tau$, $E_\text{Z}=2\,\tau$, and set the angles $\vartheta=\chi=0$.
We assumed Si/SiGe quantum dots with $\sigma=10$~nm and $d=100$~nm, and we used the material parameters $\alpha = 609$~m/s, $\Xi_d=5$~eV, $\Xi_u=9$~eV, $\rho_0=2330$~kg/m$^3$, $v_l=9150$~m/s, $v_t=5000$~m/s~\cite{Malkoc2016,Huang2014,Madelung2004,YU2010} and the transverse effective mass $m^*=0.19\,m_e$~\cite{Raith2011,Madelung2004}, for which we find $\gamma_\text{ex} = 960$~kHz. 
The value of $\beta$ is irrelevant in this case since $A_{xx}$ is independent of it for our choice or angles $\vartheta$ and $\chi$.
Fig.~\ref{fig:relax_rates}(a) shows the rate $\Gamma_1$ and Fig.~\ref{fig:relax_rates}(b) the rate $\Gamma_0$.
We see that the magnitude of the rates ranges from $\sim 10^{-12}\,\gamma_\text{ex}$ to $\sim 10^{-5} \, \gamma_\text{ex}$, which is typically much smaller than the decoherence rates due to other mechanisms, such as phonon-mediated qubit relaxation (transitions from $\ket 1$ to $\ket 0$) and dephasing caused by charge noise.
Vertical and horizontal dashed lines indicate the line cuts that we show in Figs.~\ref{fig:relax_rates}(c,d). Here we plot the leakage rates (c) as a function of $V_m$ for $\epsilon = 0$ and (d) as a function of $\epsilon$ for $V_m = -0.3\, U$. Circles and crosses present numerical results and the solid lines in (c) show the analytical results of Eqs.~(\ref{eq:rateRX1},\ref{eq:rateRX0}), which indeed agree well with the numerical results.

At the SS the qubit can also be operated electrically by tuning the tunnel barriers, without the need of leaving this point of low decoherence, as has been pointed out before~\cite{Shim2016}.
Our numerical calculations confirm that at the SS the relaxation rates between any two states in the lowest part of the spectrum (including the qubit relaxation rate $\Gamma_\text{rel}$) are strongly suppressed, not only for $t_{12}=t_{23}$ as in Fig.~\ref{fig:relax_rates}, but for any combination of tunneling energies.
The triple-dot spin qubit can be thus operated at the SS via a modulation of the tunneling amplitudes (the AEON qubit \cite{Russ2017}) while being highly insensitive to charge noise, relaxation and SO-assisted leakage.
The constant contributions of virtual valley and orbital excitation to the leakage rates (see Secs.~\ref{sec:orb} and \ref{sec:val}) is estimated to be $\sim 10^{-10} \, \gamma_\text{ex}$ for our choice of parameters and do therefore not affect these conclusions qualitatively.

\section{Dephasing at the SS}
\label{sec:dep}

Dephasing in $^{28}$Si-based triple-dot spin qubits is believed to mainly come from electric noise in the qubit's environment~\cite{Eng2015,Russ2016}.
As a first approximation, one can understand such dephasing by assuming the noise to manifest itself as fluctuations of the gate potentials, $V_i(t) = V_i + \delta V_i(t)$, that are Gaussian and have zero mean.
To leading order, the qubit frequency then acquires a time-dependence $\omega(t) = \omega + \delta\omega(t)$, with $\delta\omega(t) = \sum_i (\partial \omega / \partial V_i) \delta V_i(t)$, and a qubit prepared in the coherent superposition $\ket{\psi(0)} = \ket + = \frac{1}{\sqrt 2}(\ket 0 + \ket 1)$ will thus evolve as $\ket{\psi(t)} = \frac{1}{\sqrt 2}(\ket 0 + e^{i[\phi(t)+\delta\phi(t)]}\ket 1)$, where $\delta \phi(t) = \int^t_0 {\rm d}t'\, \delta\omega(t')$.
The noise-induced dephasing can then be characterized by investigating the expectation value $\langle e^{i \delta\phi(t)} \rangle$:
Since the fluctuations $\delta \omega (t)$, and thus the fluctuations $\delta\phi(t)$, are Gaussian, only the second cumulant in the expansion of $\langle e^{i \delta\phi(t)} \rangle$ is non-zero, resulting in low-frequency noise in an exponential decay of the coherent qubit oscillations $\sim e^{-t^2/T_\varphi^2}$, where the exact form of the dephasing time $T_\varphi$ depends on the detailed noise spectrum~\cite{Russ2015,Ithier2005,Makhlin2004,Russ2016}.

Exactly at the SS, all first-order derivatives $\partial \omega/\partial V_i$ vanish, and therefore this type of dephasing is highly suppressed.
To understand the remaining charge-noise-induced dephasing at the SS, one could thus try to use the same approach but now focus on the next order, $\delta\omega(t) = \sum_{i,j} (\partial^2 \omega / \partial V_i\partial V_j) \delta V_i(t)\delta V_j(t)$.
In this case, however, the fluctuations $\delta\omega(t)$ [and thus $\delta\phi(t)$] are no longer Gaussian and one would thus have to include all cumulants in the expansion of $\langle e^{i \delta\phi(t)} \rangle$~\cite{Cywinski2014}.
Therefore it is more convenient to investigate the explicit time evolution of the qubit~\cite{Laird2010,Taylor2013,Russ2017,Malinowski2017}.
Assuming for simplicity quasistatic fluctuations~\cite{Russ2015,Ithier2005,Makhlin2004}, we can evaluate the time-dependent probability $P(t) = | \langle{ {+}|\psi(t)} \rangle|^2$ to find the qubit in the state $\ket +$ and average this probability over the fluctuations $\delta V_i$~\cite{Koppens2007}.
To good approximation we then find (see Appendix~\ref{appB} for details)
\begin{align} \label{eq:prob}
	\langle P(t) \rangle = {}&{} \frac{1}{2} + \frac{\cos\left( \omega t - \arctan [t/T_\varphi]\right)}{2\sqrt{ 1+ t^2/T^2_\varphi }},
\end{align}
where $T_\varphi = \hbar(U-U_c)^3/4\tau^2\xi^2$ is the dephasing time, with $\xi^2 = \langle (\delta V_i)^2 \rangle$ the variance of the fluctuations.
At the SS the leading-order contribution of the charge noise to dephasing thus results in (\textit{i}) a time-dependent phase shift in the qubit oscillations, which goes to $-\pi/2$ for $t \gtrsim T_\varphi$, and (\textit{ii}) a decay of the coherent oscillations with a power-law behavior, $\sim T_\varphi/t$ for large times, in contrast with the exponential decay $\sim e^{-t^2/T_\varphi^2}$ one finds away from the SS, whenever $\sum_i (\partial \omega / \partial V_i) \delta V_i(t) \gg \sum_{i,j} (\partial^2 \omega / \partial V_i\partial V_j) \delta V_i(t)\delta V_j(t)$~\cite{Russ2016}.
One can also use a detailed cumulant-expansion approach to describe quadratic coupling to Gaussian noise, which leads to the same long-time behavior as we found here~\cite{Cywinski2014}.

\begin{figure}[t!]
	\includegraphics[width=\linewidth]{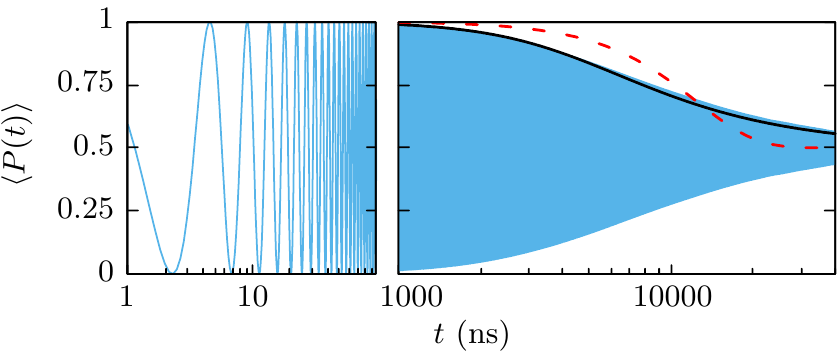}
	\caption{Numerically calculated time-dependent return probability $| \langle{ {+}|\psi(t)} \rangle|^2$ after initializing in $\ket +$, averaged over $10^5$ different sets of $\delta V_{1,2,3}$ taken from a normal distribution with $\xi = 5~\mu$eV (thin blue line).
	The thick black line shows the envelope function of the oscillations as predicted by Eq.~(\ref{eq:prob}) and the dashed red line shows the best fit obtainable assuming an exponential envelope of the form $\frac{1}{2} + \frac{1}{2}e^{-t^2/\tilde T^2_\varphi}$.\label{fig:deph}}
\end{figure}

We can also calculate the averaged probability $\langle P(t)\rangle$ numerically, again assuming quasistatic charge noise.
Using the same parameters as before, we tune the Hamiltonian (\ref{eq:hubb}) to the SS ($\epsilon = 0$, $V_m = -0.3\, U$) but then add random offsets $\delta V_{1,2,3}$, taken from a normal distribution with $\xi = 5~\mu$eV~\cite{Martins2016}.
We diagonalize the resulting Hamitonian, identify the two qubit states $\ket 1$ and $\ket 0$, and create an initial state $\ket + = \frac{1}{\sqrt 2}(\ket 0 + \ket 1)$.
We then evaluate numerically the time-dependent qubit state  $\ket{\psi(t)} = \exp\{-\frac{i}{\hbar} \hat H t\} \ket +$, and from this we can calculate $P(t)$ for the specific set of $\delta V_i$ chosen.
This procedure is repeated $10^5$ times, and the resulting average $\langle P(t) \rangle$ is shown by the blue curve in Fig.~\ref{fig:deph}.
As expected, we see an oscillating probability that decays to $1/2$ over time.
The black solid line shows the envelope function of the decaying oscillations, as given by Eq.~(\ref{eq:prob}), where we have $T_\varphi \approx 4~\mu$s for our choice of parameters.
We see that the power-law decay predicted by (\ref{eq:prob}) matches the numerical results very well.
For comparison we included a best fit of the form $\frac{1}{2} + \frac{1}{2}e^{-t^2/\tilde T^2_\varphi}$ (red dashed line), which yields $\tilde T_\varphi \approx 12.4~\mu$s but indeed shows a much worse agreement with our numerical results than the power-law from (\ref{eq:prob}).

\section{Conclusions}
\label{sec:conc}

Leakage out of the qubit subspace in XO qubits hosted in $^{28}$Si-based triple quantum dots is caused mainly by SO interaction via virtual spin-flip tunneling.
Together with spin-conserving phonon emission, this results in tuning-dependent leakage rates that scale as $\Gamma_1\propto E_\text{Z}^7$ and $\Gamma_0\propto E_\text{Z}^5$ and are strongly reduced at the SS, where the qubit is minimally sensitive to charge noise as well.
We found that the other two mechanisms of leakage we investigated, virtual orbital and valley excitation, result in much smaller relaxation rates, scaling as $\Gamma_1 \propto E_\text{Z}^9$ and $\Gamma_0 \propto E_\text{Z}^{11}$; they are constant throughout the entire (1,1,1) charge region, thus becoming the most relevant mechanism of leakage only at the SS.
Further, we showed that also (spin-conserving) qubit relaxation, enabled by electron-phonon coupling, is minimal at the SS, making this an ideal operation point in many respects.

We also investigated the residual effects of charge noise at the SS, which are most likely the dominating source of pure dephasing at that point.
We found that slow electric fluctuations result in dephasing that makes coherent qubit oscillations decay as $\propto 1/t$, in analogy to Ref.~\cite{Koppens2007,Cywinski2014}; this in contrast with the exponential decay that dominates elsewhere in the (1,1,1) charge region.
%

This work was partly supported by the Research Council of Norway through its Centers of Excellence funding scheme, project number 262633, QuSpin.

\appendix

\section{Detailed analytic results}
\label{appA}

The dimensionless functions used in the analytic results presented in Sec.~II read explicitly
\begin{multline}
	f_{1}^\text{ex}(x) =  \frac{4}{5} + \frac{1}{16 x^5} \left[
	128x(x^2-9)\cos x \right. \\  - 2(4x^2-9)(x \cos 2x +64 \sin x)  \\ \left.+ (16x^2-9)\sin 2x  \right],
\end{multline}
\begin{multline}
	f_{0}^\text{ex}(x) =  \frac{4}{15} + \frac{1}{16x^5} \left[
	2x(4x^2-9)\cos 2x \right. \\ \left. - (16x^2-9)\sin 2x \right],
\end{multline}
\begin{multline}
	f_{1}^\text{orb}(x) = 
	\frac{8}{105}(a_{xx}^2 + a_{yx}^2) \\
	+ \frac{1}{32 x^6} \big[ 3 a_{yx}^2 (8x^2-15) \\ + 2 a_{xx}^2(8x^4-78 x^2+135)\big] \cos 2x \\ 
	+ \frac{1}{64 x^7} \big[ a_{yx}^2(16x^4-84x^2+45) \\ - 2 a_{xx}^2(64x^4-258x^2+135) \big] \sin 2x, \label{eq:orb1}
\end{multline}
\begin{multline}
	f_{0}^\text{orb}(x) = 
	\frac{8}{35}(a_{xx}^2+a_{yx}^2) \\
      + \frac{1}{64 x^7}\Big\{ 512 x \big[ 3 a_{yx}^2 (2 x^2-15 ) \\ + a_{xx}^2 (x^4 -39 x^2+270) \big] \cos x \\
	-2 x \big[ 3 a_{yx}^2 (8x^2-15) \\ + 2 a_{xx}^2 (8x^4 -78 x^2 + 135) \big] \cos 2x \\
	+512 \big[ a_{yx}^2 (x^4 - 21x^2 + 45) \\ - a_{xx}^2 (8x^4 -129 x^2 + 270) \big] \sin x \\
	- \big[a_{yx}^2 (16x^4 -84x^2 +45) \\ - 2 a_{xx}^2 (64 x^4 - 258x^2 + 135) \big] \sin 2x \Big\}, \label{eq:orb0}
\end{multline}
\begin{multline}
	f_{1}^\text{val}(x) = 
	\frac{16}{105}
	+ \frac{1}{64 x^7}
	\big[ 2x (16x^4-132 x^2+225) \cos 2x \\
	-(112x^4-432x^2+225) \big] \sin 2x,
\end{multline}
\begin{multline}
	f_{0}^\text{val}(x) = 
	\frac{16}{35}
	+ \frac{8}{x^7}
	\big[ x (x^4 -33 x^2+225) \cos x \\
	- (7x^4 -108 x^2 + 225) \sin x \big] \\
	-\frac{1}{64x^7}
	\big[ 2 x (16 x^4 - 132 x^2 + 225)\cos 2x \\ - (112 x^4 -432 x^2 + 225)\sin 2x \big],
\end{multline}
where the spin-orbit velocities in (\ref{eq:orb1},\ref{eq:orb0}) are rescaled with the transverse phonon velocity, $a_{xx,yx} \equiv A_{xx,yx}/v_t$.

\section{Charge noise and dephasing at the sweet spot}
\label{appB}

In the absence of significant hyperfine interaction, the main source of decoherence for exchange-based spin qubits is believed to be (low-frequency) charge noise on the gate electrodes~\cite{Russ2017,Zhang2018,Martins2016}.
Such noise results in fluctuations of the onsite potentials as used in the Hamiltonian $\hat H$ in Eq.~(\ref{eq:hubb}),
\begin{align}
	V_i(t) = V_i + \delta V_i (t).
\end{align}
This causes the projected qubit Hamiltonian to fluctuate as well,
\begin{align}
	\hat H_\text{qubit} = \frac{\hbar}{2} \big[ \omega_0 + \delta\omega_z(t) \big] \hat \sigma_z + \frac{\hbar}{2}\delta\omega_x(t)\hat\sigma_x.
\end{align}
We focus on pure dephasing in this qubit basis, i.e., we investigate how the phase of the qubit gets randomized through the fluctuations in the qubit splitting $\delta\omega_z(t)$.
To this end, we consider the system to be prepared in the state $\ket{+} = \frac{1}{\sqrt{2}} (\ket 0 + \ket 1)$ at $t=0$.
After some time $t$, the system evolved into the state $\ket{\psi(t)} = \frac{1}{\sqrt{2}} (\ket 0 + e^{i\phi(t)}\ket 1)$, where $\phi(t) = \omega_0 t + \delta\phi(t)$, the unknown part of the phase being $\delta\phi(t) = \int_0^t \delta\omega_z(t') \d{t'}$.

The expectation value of this random component of the phase can be found by evaluating $\log \langle e^{i\phi(t)} \rangle$. For simplicity we will assume quasistatic (time-independent during each individual time-evolution) Gaussian noise in the $V_i$:
\begin{align}
\delta\phi(t) = {}&{} t\sum_{i=1}^3 \frac{\partial \omega_z}{\partial V_i} \delta V_i +\frac{t}{2} \sum_{i,j}\frac{\partial^2 \omega_z}{\partial V_i \partial V_j} \delta V_i \delta V_j +  \mathcal{O}(\delta V^3).\label{eq:expphase}
\end{align}
Usually, one then focuses on the leading (first-order) term, which is linear in the fluctuations $\delta V_i$.
This makes $\delta \omega_z$ also a Gaussian variable, and then one can do a cumulant expansion of the logarithm,
\begin{align}
\log \langle e^{i\delta\phi(t)} \rangle = {}&{} \sum_{n=1}^\infty \kappa_n \frac{(i t)^n}{n!},
\end{align}
with $\kappa_n$ the $n$-th cumulant of the distribution of $\delta\omega_z$, and use that for Gaussian variables with zero mean only the second cumulant $\kappa_2=\sum_i (\partial \omega_z/\partial V_i)^2\langle \delta V_i^2 \rangle$ is non-zero.
This yields the familiar result $\log \langle e^{i\phi(t)} \rangle = -\frac{1}{2} t^2 \kappa_2$, from which one can extract an approximate dephasing time.

At the sweet spot, however, where we expect this dephasing time to be maximal, the first derivative of $\omega_z$ vanishes (per definition~\cite{Shim2016,Russ2016}) and one has to use the next (second-)order term in the series expansion of the phase (\ref{eq:expphase}).
A subtle point, sometimes overlooked, is that, although the fluctuations $\delta V_i$ are Gaussian, the product $\delta V_i \delta V_j$ of two Gaussian random variables is not Gaussian anymore.
This implies that the cumulant expansion has many more non-zero terms that become relevant at long times, causing $\log \langle e^{i\phi(t)} \rangle \neq -\frac{1}{2} t^2 \kappa_2$~\cite{Cywinski2014}.
One can only use such an equality as long as $\frac{\partial \omega_z}{\partial V_i} \gg  \frac{\partial^2 \omega_z}{\partial V_i \partial V_j}$, a condition that is not satisfied at the sweet spot.

To extend the analysis to the sweet spot we focus on the Schr\"odinger equation resulting from the effective qubit Hamiltonian instead.
We will consider only a diagonal Hamiltonian,
\begin{align}
i\hbar\frac{\partial}{\partial t} \psi(t) = \frac{\hbar}{2} (\omega_z + \delta \omega_z) \sigma^z \psi(t)
\end{align}
and use again the initial condition $\left| \psi(t=0) \right\rangle = \ket + = \frac{1}{\sqrt 2}(| 0 \rangle + | 1 \rangle)$, with $|0\rangle$ and $|1\rangle$ two eigenvalues of the qubit Hamiltonian for $\delta\omega_z = 0$.
In this case, the probability of finding the qubit in the initial state $\left| {+} \right\rangle$ after time $t$ is
\begin{align} \label{eq:prob_full}
P = \left| \left\langle {+} \middle| \psi(t) \right\rangle \right|^2 = {}&{} 
\cos\left(\frac{t [\omega_z + \delta\omega_z]}{2 }\right)^2.
\end{align}
For the exchange-only qubit at the sweet spot, the fluctuation $\delta\omega_z$ is given by the second-order term
\begin{align}
\delta\omega_z = {}&{} \frac{\tau^2}{(U-U_c)^3}\left[ (\delta V_2 - \delta V_1)^2 + (\delta V_2 - \delta V_3)^2 \right],
\end{align}
where we again have set $t_{12}=t_{23}\equiv\tau$.

In order to average over the fluctuations, we define two variables $x_1 = \delta V_2 - \delta V_1$ and $x_2 = \delta V_2 - \delta V_3$ that we will consider independent, for simplicity.
These variables have mean 0 and standard deviation $\sqrt 2 \xi$ (with $\xi$ the standard deviation of the original variables $\delta V_i$), and can be combined into one $\chi^2$-distributed random variable $y = \frac{x_1^2}{2\xi^2} + \frac{x_2^2}{2\xi^2}$.
With this the probability becomes
\begin{align}
P = {}&{} \frac{1}{2}\int_0^\infty \!\!\! \d y e^{-y/2} \cos^2\left[ \frac{t}{2 \hbar} \left( \frac{2 \tau^2 }{U-U_c} + \frac{2 \sigma^2 \tau^2}{(U-U_c)^3} y \right) \right].
\end{align}
This integral can be solved analytically, yielding
\begin{align} 
	P = {}&{} \frac{1}{2} + \frac{\cos\left( t \frac{2\tau^2}{\hbar (U-U_c)} - \arctan \big[ t \frac{4 \tau^2 \xi^2}{\hbar(U-U_c)^3} \big]\right)}{2\sqrt{ 1+ t^2 \frac{16 \tau^4\xi^4}{\hbar^2(U-U_c)^6} }},
\end{align}
We see that, as expected, the probability oscillates with a frequency $2\tau^2/\hbar(U-U_c)$ (while also gradually acquiring a phase shift that goes to $-\pi/2$ for $t \to \infty$).
The amplitude of the oscillations decays within the envelope function
\begin{align}
P_\text{env} = {}&{} \frac{1}{2} + \frac{1}{2 \sqrt{1 + t^2 \frac{16 \tau^4 \xi^4}{\hbar^2 (U-U_c)^6}}},
\end{align}
which, for long times predicts a decay $\propto T_\varphi / t$ with a dephasing time of $T_\varphi = \hbar (U-U_c)^3 / 4 \tau^2\xi^2$.
This is in contrast with the exponential decay $\propto e^{-t^2/T_\varphi^2}$ that is predicted by the ``cumulant expansion method''~\cite{Russ2015,Russ2016}.
A simple simulation of the time evolution of the state $\ket +$ under the action of quasistatic random noise at the sweet spot corroborates this result (see Sec. IV).

%

\end{document}